# Impact of Ultrasound on the Motion of Compact Particles and Acousto-responsive Microgels

Sebastian Stock, Regine von Klitzing, and Amin Rahimzadeh*

Soft Matter at Interfaces, Institute for condensed matter physics, Technische Universität Darmstadt, Hochschulstraße 8, 64289 Darmstadt, Germany

## Abstract

In this study, we investigate dynamic light scattering (DLS) from both randomly diffusing silica particles and acousto-responsive microgels in aqueous dispersions under ultrasonic vibration. Employing high-frequency ultrasound (US) with low amplitude ensures that the polymers remain intact without damage. We derive theoretical expressions for the homodyne autocorrelation function, incorporating the US term alongside the diffusion term. Subsequently, we successfully combine US with a conventional DLS system to experimentally characterize compact silica particles and microgels under the influence of US. Our model allows us to extract essential parameters, including particle size, frequency, and amplitude of particle vibration, based on the correlation function of the scattered light intensity. The studies involving non-responsive silica particles demonstrate that US does not disrupt size determination, establishing them as suitable reference systems. Microgels show the same swelling/shrinking behavior as that induced by temperature, but with significantly faster kinetics. The findings of this study have potential applications in various industrial and biomedical fields that benefit from the characterization of macromolecules subjected to US.

## I. Introduction

Dynamic light scattering (DLS) is commonly used for the characterization of particles and molecules in solutions/dispersions, such as determining their size, size distribution as well as conformational changes. DLS offers to measure sub-micrometer (from a few nanometers to one micrometer) particles accurately. When a laser beam impinges the particles inside a liquid sample, the particles scatter light in all directions. A detector at a certain location detects a fraction of the scattered light and measures it as intensity fluctuations over time. In a conventional DLS system, these fluctuations are caused by interference between scattered light from an ensemble of particles moving due to Brownian motion. The intensity fluctuations can be analyzed by calculating the time-dependent correlation of the signal with itself in different time lags which is called the autocorrelation analysis technique. The particle size can be determined by the decay rate of the autocorrelation function according to the Stokes-Einstein relation [1,2]. According to the Brownian motion theory, smaller particles diffuse faster and thus exhibit shorter correlation times [3].

In many cases, information about the mechanical, conformational and electrical properties of particles or molecules, including their size, shape, and surface charges are needed under the influence of external forces or fields. Therefore, *dynamic light scattering in external fields* has many applications in biophysics, materials science, and chemical engineering [4]. In those systems– e.g., particles in alternating electric fields [5,6], in thermal in-homogeneities [7,8], or directional flows [9]– colloidal particles undergo another movement in addition to their original random motion. These additional movements create complicated intensity fluctuations from which the autocorrelation function (ACF)– with the conventional fitting parameters– does not impart the diffusion coefficient and subsequently the correct particle size anymore [10]. Therefore, one should modify the ACF fitting parameter in order to include the additional translational motion and to distinguish it from the pure Brownian diffusion. For instance, in the case of DLS measurements of colloidal particles in a flowing condition, researchers have modified the ACF so that they could obtain the correct particle size as well as the flowing velocity [9,11–13]. One of the external fields that have drawn attention in recent decades is US which is employed to manipulate the physical or chemical properties of particles and molecules with applications in drug delivery [14], catalysis [15] and materials synthesis [16]. Recently, we showed that high-frequency US, in its non-destructive condition (low amplitude), can be used as a stimulus to induce a phase transition in solutions of linear poly(N-Isopropylacrylamide)(PNIPAM) [17,18]. The dehydration of PNIPAM which usually triggered by increasing the temperature above the lower critical solution temperature (LCST), is then induced by US. In the case of linear PNIPAM, the phase transition is detectable by an onset of turbidity. PNIPAM microgels are cross-linked polymer networks that upon a stimulus shrink and reduce in size having a promising application in drug delivery systems. They are also temperature sensitive and their dehydration might be also induced by US. In order to monitor the microgel size (or any other responsive particles) subjected to US, one has to modify the DLS system and analyze the resulting ACF of standard particles under the influence of ultrasonic waves as a reference system. Beside volume phase transition (VPT), ultrasound might lead to other disturbances like fluctuation of the microgels trajectory or acoustic streaming, which might induce an apparent size change of the microgels. In order to separate these effects, in this work, first we determine the size of solid particles using DLS subjected to US, assuming that they are shape invariant in US. In the first step, we derive the homodyne correlation function for vibrating particles based on the continuity equation in the work done by Berne and Pecora [19]. Then, we compare the results to a representative system of silica nanospheres subjected to US. This system is a simple combination of a conventional DLS setup with an US component that can establish a ground for more complicated experimental systems.

Based on the evaluated results of the reference system, we conduct an experiment using a sample of PNIPAM microgels to assess their US-induced VPT using their changes in size. The microgel diameter over ultrasonic actuation time is compared with its diameter with the increase in temperature.

## II. Deriving equations

Here we show the principal derivation of the intensity autocorrelation function of the scattered light from an ensemble of particles undergoing Brownian diffusional motion as well as an external-induced flow motion with the velocity $\boldsymbol{v}$. The particle concentration at point $\boldsymbol{r}$ and time $t$ is defined by $c(\boldsymbol{r},t)$. The continuity equation which describes how the particles flow and diffuse (having a diffusion coefficient D) in the system can be written as:

$$\frac{\partial c}{\partial t} + \nabla \cdot (\boldsymbol{v}c) = D\nabla^2 c \tag{1}$$

According to Berne and Pecora [19], whose notation we adopted in this study, it is reasonable to assume that the *probability distribution function*, $G_s(\boldsymbol{r},t)$, satisfies the same equation. Therefore, we have:

$$\frac{\partial G_s}{\partial t} + \nabla \cdot (\boldsymbol{v}G_s) = D\nabla^2 G_s \tag{2}$$

We consider the *characteristic function of distribution*, $F_s(\boldsymbol{q},t)$ as the Fourier transform of $G_s$, based on the following definitions:

$$F_s(\boldsymbol{q},t) = \int G_s(\boldsymbol{r},t)\exp(i\boldsymbol{q}\cdot\boldsymbol{r})\,d^3r, \tag{3}$$

$$G_s(\boldsymbol{r},t) = (2\pi)^{-3}\int F_s(\boldsymbol{q},t)\exp(-i\boldsymbol{q}\cdot\boldsymbol{r})\,d^3q \tag{4}$$

where $q = \frac{4\pi n \sin\left(\frac{\theta}{2}\right)}{\lambda_l}$ is the scattering wave vector with a wavelength of $\lambda_l$ and scattering angle of $\theta$ and the medium refractive index of $n$. In case the system is subjected to US waves having a wave vector of $|\boldsymbol{k}| = 2\pi/\lambda_u$ ($\lambda_u$ is the US wavelength) and angular frequency of $\omega$, the velocity of the fluid at point $\boldsymbol{r}$ and time $t$, having a mean value $\boldsymbol{v_0}$ can be written as:

$$\boldsymbol{v}(\boldsymbol{r},t) = \boldsymbol{v_0}\exp(i\boldsymbol{k}\cdot\boldsymbol{r} - i\omega t). \tag{5}$$

By taking the spatial Fourier transform from eq. (2), the first term on the left-hand side and the term on the right-hand side yield to $\frac{\partial F_s(\boldsymbol{q},t)}{\partial t}$ and $-Dq^2F_s(\boldsymbol{q},t)$, respectively. The second term on the left-hand side can be written as (Ft: Fourier transform):

$$Ft[\boldsymbol{v}\cdot\nabla G + G\nabla\cdot\boldsymbol{v}] = \int \boldsymbol{v}\cdot\frac{\partial G_s}{\partial r}\exp(i\boldsymbol{q}\cdot\boldsymbol{r})\, d^3r + \int G_s\frac{\partial \boldsymbol{v}}{\partial r}\exp(i\boldsymbol{q}\cdot\boldsymbol{r})\, d^3r. \tag{6}$$

Using the eq. (5), the eq. (6) yields to:

$$\boldsymbol{v_0}\exp(-i\omega t)\int\frac{\partial G_s}{\partial r}\mathrm{e}^{i(\boldsymbol{q}+\boldsymbol{k})\cdot\boldsymbol{r}}d^3r + i\boldsymbol{k}\cdot\boldsymbol{v_0}\exp(-i\omega t)\int G_s\mathrm{e}^{i(\boldsymbol{q}+\boldsymbol{k})\cdot\boldsymbol{r}}d^3r, \tag{7}$$

where $\int\frac{\partial G_s}{\partial r}\mathrm{e}^{i(\boldsymbol{q}+\boldsymbol{k})\cdot\boldsymbol{r}}d^3r = -iq\int G_s\mathrm{e}^{i(\boldsymbol{q}+\boldsymbol{k})\cdot\boldsymbol{r}}d^3r$. The reason is that taking a derivative from eq. (4), leads to:

$$\frac{\partial}{\partial r}G_s(\boldsymbol{r},t) = -iq(2\pi)^{-3}\int F_s(\boldsymbol{q},t)\exp(-i\boldsymbol{q}\cdot\boldsymbol{r})\, d^3q = -iqG_s. \tag{8}$$

Therefore, eq. (7) can be written as:

$$\begin{aligned}-i\boldsymbol{q}\cdot\boldsymbol{v_0}\exp(-i\omega t)\int G_s\mathrm{e}^{i(\boldsymbol{q}+\boldsymbol{k})\cdot\boldsymbol{r}}d^3r + i\boldsymbol{k}\cdot\boldsymbol{v_0}\exp(-i\omega t)\int G_s\mathrm{e}^{i(\boldsymbol{q}+\boldsymbol{k})\cdot\boldsymbol{r}}d^3r \\ = -i(\boldsymbol{q}-\boldsymbol{k})\cdot\boldsymbol{v_0}\exp(-i\omega t)\, F_s(\boldsymbol{q}+\boldsymbol{k},t).\end{aligned} \tag{9}$$

Finally, the spatial Fourier transform of eq. (2) leads to:

$$\frac{\partial F_s(\boldsymbol{q},t)}{\partial t} - i(\boldsymbol{q}-\boldsymbol{k})\cdot\boldsymbol{v_0}\exp(-i\omega t)\, F_s(\boldsymbol{q}+\boldsymbol{k},t) = -Dq^2F_s(\boldsymbol{q},t). \tag{10}$$

In the case of a uniform flow with the velocity of $\boldsymbol{v_0}$ instead of ultrasonic vibration (i.e., $k=\omega=0$), eq. (10) reduces to the same equation derived by Berne and Pecora [19] as:

$$\frac{\partial F_s(\boldsymbol{q},t)}{\partial t} - i\boldsymbol{q}\cdot\boldsymbol{v_0}F_s(\boldsymbol{q},t) = -Dq^2F_s(\boldsymbol{q},t). \tag{11}$$

With the initial condition of $F_s(\boldsymbol{q},0)=1$, we have $F_s(\boldsymbol{q},t)=\exp(-Dq^2t)\exp(i\boldsymbol{q}\cdot\boldsymbol{v_0}t)$.

At sufficiently low US frequencies (i.e., less than 1 GHz) in water, as the medium, $q \gg k$. Therefore, the eq. (10) can be simplified as:

$$\frac{\partial F_s(\boldsymbol{q},t)}{\partial t} - i\boldsymbol{q}\cdot\boldsymbol{v_0}\exp(-i\omega t)\, F_s(\boldsymbol{q},t) = -Dq^2F_s(\boldsymbol{q},t). \tag{12}$$

By solving the eq. (12) with the initial condition of $F_s(\boldsymbol{q},0)=1$, $F_s$ can be obtained as:

$$F_s(\boldsymbol{q},t) = \exp\{-Dq^2t\}\exp\{-\frac{\boldsymbol{q}\cdot\boldsymbol{v_0}}{\omega}\exp(-i\omega t)\}. \tag{13}$$

Assume that $\boldsymbol{v_0}=\boldsymbol{r_0}\omega$, where $r_0$ is the amplitude of vibration, eq. (13) can be rewritten as:

$$F_s(\boldsymbol{q},t) = \exp\{-Dq^2t\}\exp\{-\boldsymbol{q}\cdot\boldsymbol{r_0}\exp(-i\omega t)\}. \tag{14}$$

From Berne and Pecora [19], the homodyne correlation function can be obtained as:

$$\begin{aligned}F_2(\boldsymbol{q},t) &= \langle N\rangle^2[1+|F_s(\boldsymbol{q},t)^2|] + \langle\delta N(0)\delta N(t)\rangle \\ &= \langle N\rangle^2[1+Re(\exp\{-Dq^2t\}\exp\{-2\boldsymbol{q}\cdot\boldsymbol{r_0}\exp(-i\omega t)\}] + \langle\delta N(0)\delta N(t)\rangle\end{aligned} \tag{15}$$

Therefore, in the homodyne experiment, the autocorrelation function for the time lag, $\tau$, can be written as:

$$g_2(\tau) = A[1+B\exp\{-2Dq^2\tau\}\exp\{-2qr_0\cos(\omega\tau)\}]. \tag{16}$$

The B is called intercept, related to light-collection efficiency, and can be omitted by normalizing the autocorrelation function [20]. Therefore, the normalized autocorrelation function (NACF) can be written as:

$$NACF = \frac{g_2(\tau) - A}{B} = \exp\{-2Dq^2\tau\}\exp\{-2qr_0\cos(\omega\tau)\}. \quad (17)$$

## III. Experimental

The experimental setup, as schematically shown in Figure 1a, comprises a HeNe gas laser (HNL150L, Thorlabs GmbH, Germany, $\lambda = 633\ nm$), a photodetector (APD130A2/M, Thorlabs GmbH, Germany) detecting the scattered light at an angle of 90°, data acquisition system (BNC-2110, National Instruments, USA, sampling rate = 1 MS/s) connected to a computer. Piezoelectric transducers (STEMINC-PIEZO, Davenport, IA, USA) with different resonance frequencies (40 kHz, 255kHz, 780 kHz, 2.34 MHz, and 5.4 MHz) were used and attached to the glass cuvette (inside dimensions of 10×10×40 $mm^3$) using a two-component latex glue (UHU Endfast Plus300, Bühl, Germany). The RF signal was generated using a function generator (SDG1062X, SIGLENT, Shenzhen, China) and was amplified by an RF amplifier (VBA100-30, Vectawave, UK). We used silica nanospheres (nanoComposix, CA, USA) in different diameters (80, 200, 500, and 1000 nm) as colloidal dispersions diluted in milli-Q water. PNIPAM microgel particles with 5 mol% cross-linker content (BIS) were synthesized using precipitation polymerization. For detailed information see our previous works [21,22]. A commercially available DLS system (LS Instrument, Switzerland) was used to measure the hydrodynamic diameter of PNIPAM microgels due to changes in temperature.

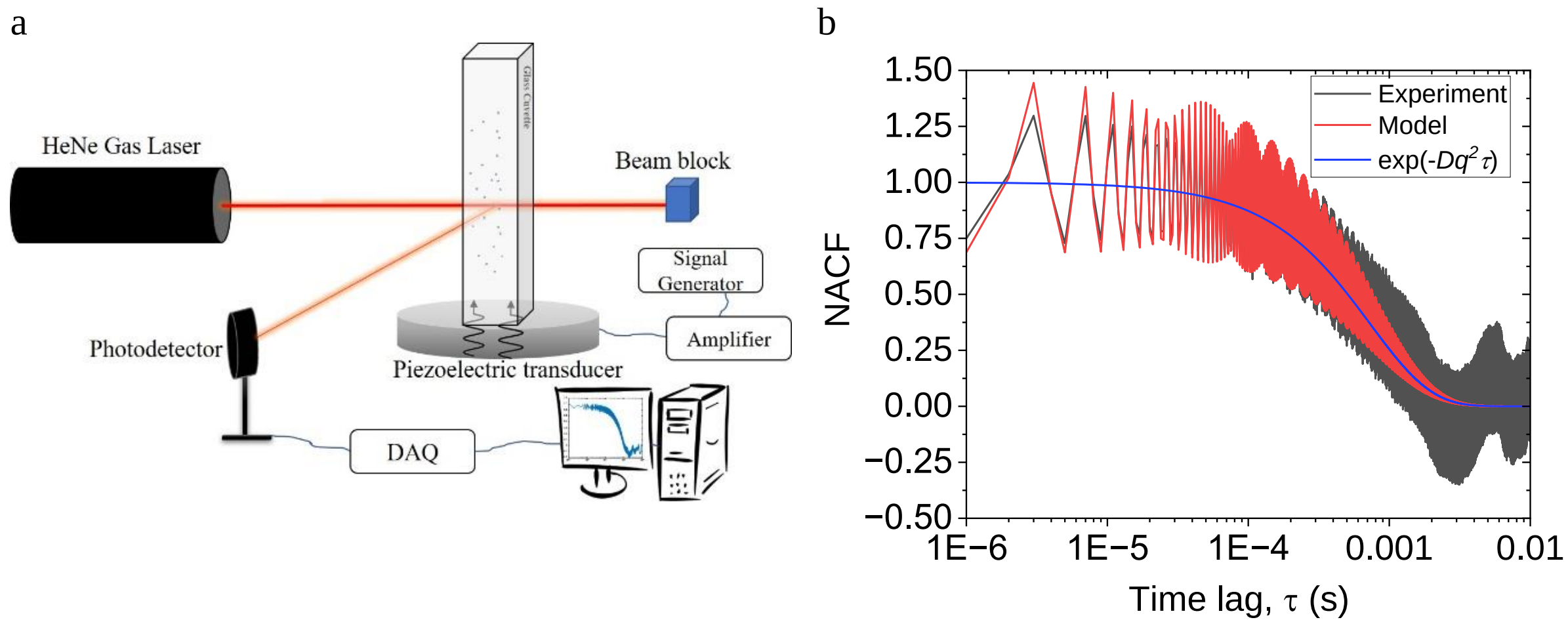

Figure 1. (a) Schematic of the experimental apparatus comprising a DLS setup and US system connected to the sample via a piezoelectric transducer and glass cuvette. (b) Normalized autocorrelation function (NACF) for 200 nm silica particles subjected to 255 kHz and 300 mV US.

## IV. Results and discussion

According to Eq. (17), the photon intensity correlation function exhibits an oscillatory behavior. Such behavior previously was observed for particles in alternating electric fields at very low frequencies [4]. From the modified NACF, the diffusion coefficient due to Brownian motion and hence the particle size can be acquired independently of the US effect. Moreover, the oscillatory component in Eq. (17) specifies both the frequency and amplitude of the particle's vibration. The experimental results are analyzed using our MATLAB code based on linear cumulant analysis and are compared with the model. The NACF in both the experiment and model (shown in Figure 1b) for 200 nm silica particles subjected to 255 kHz US shows a good agreement. However, the oscillations damp gradually in the model while the experiments show a fixed amplitude of oscillations. Nevertheless, NACF at 255 kHz gives us valuable information about the frequency and amplitude of vibrations (Figure 2). The frequency of these oscillations– which is shown clearer using the linear time scale in the graph inset– is equal to the input frequency of the US (Figure 2a). The curve without US presents a mean value around which the data recorded in the presence of US fluctuates. The amplitude of oscillations increases by increasing the input voltage although the frequency and phase remain constant. Figure 2b shows how the particles with different sizes behave under the same US properties. Particles of smaller sizes, such as those with 80 nm diameter, demonstrate significantly greater oscillation amplitude compared to their larger counterparts, like those with a size of 1000 nm. However, despite the amplitude difference, both small and large particles exhibit the same frequency and phase of oscillation. The extracted amplitude of vibration based on our model in Eq. (17) and Figure 2b is shown in Table 1 for all particles. Despite that the order of magnitude of the amplitudes in Table 1 seems reasonable, unfortunately there is no other experimental method to confirm the values.

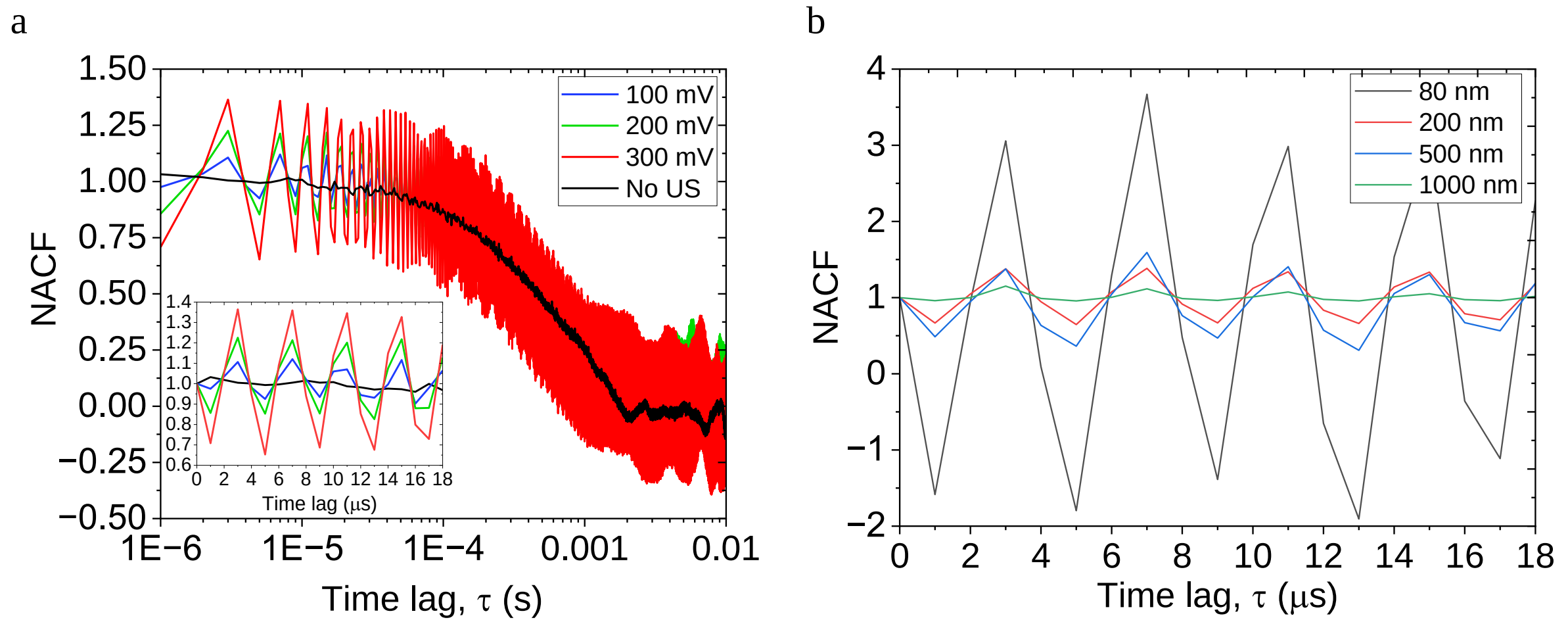

Figure 2. Normalized autocorrelation function (NACF) of (a) 200 nm particles at different input voltages and (b) particles with different diameters at 300 mV input voltage. In both cases, the frequency of US is set to 255 kHz.

Based on our theory and experimental results, we can assert that the estimation of particle size can be accurately determined regardless of the frequency and amplitude of the US used (Figure 3a). The frequency of US, which is identical to that of the particles, can be extracted from NACF provided that the data acquisition system has enough sampling rate. Figure 3b shows that at frequencies higher than half of the sampling rate (which is 500 kHz) the oscillations cannot be captured in the correct frequency complying with the Nyquist–Shannon sampling theorem [23]. Now, the question is how acoustic streaming does not interfere with the DLS results. Generally, when a liquid is subjected to US, a flow field develops due to the absorption of ultrasonic waves by the liquid viscosity. This flow may influence the diffusion of particles by increasing the translational velocity of particles leading to a wrong size estimation. However, in our experiments, the diffusion time scales for the range of particle sizes investigated are considerably shorter than the time scales associated with acoustic streaming. As a result, acoustic streaming does not influence the extracted diffusion coefficient. From the work by Leung et al. [11] one can realize that when there is a uniform flow having a low velocity (less than 1 cm/s), the particle size can be obtained with an acceptable error using the conventional intensity correlation functions. Acoustic streaming often has a velocity range of less than 1 mm/s [24]. That is why for the sub-micrometer particles, the diffusion is fast enough that the effect of acoustic streaming is negligible.

Taking the advantage of the analyzed reference system, we studied the US-induced VPT of PNIPAM microgels. As it is shown in Figure 4a, the hydrodynamic diameter of microgels (from 690 nm at swollen state) decrease (to 232 nm at collapsed state) due to imposing US over time. This is the same size as achieved with increasing temperature above the VPTT (Figure 4b), but with much faster kinetics. The reduction of microgel size after 10 seconds of actuation implies fast dehydration upon imposing US. However, the larger error bars during the transient size-change may arise due to two factors: the inhomogeneity of the particle sizes (collapsed and swollen microgels) within the measured spot and the measurement duration (about 10 seconds), which falls within the continuous actuation period.

Table 1. Vibration amplitude of nanoparticles subjected to 255 kHz and 300mV US extracted from NACF.

| **Particle diameter (nm)** | **80** | **200** | **500** | **1000** |
|---|---|---|---|---|
| **Vibration $r_0$ (nm)** | 61.95±8.35 | 10.7±1.6 | 14.06±2.01 | 4.68±0.66 |

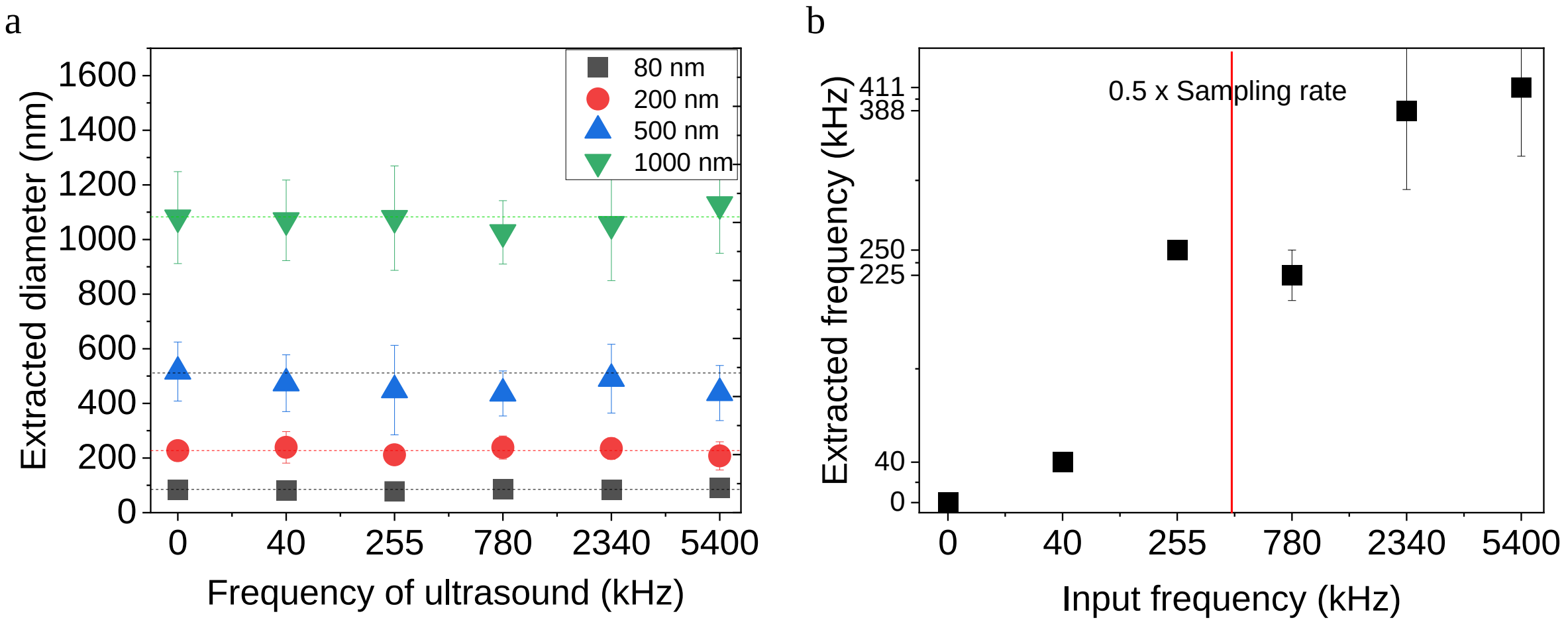

Figure 3. Extracted (a) diameter of silica particles and (b) frequency of particles vibration from NACF.

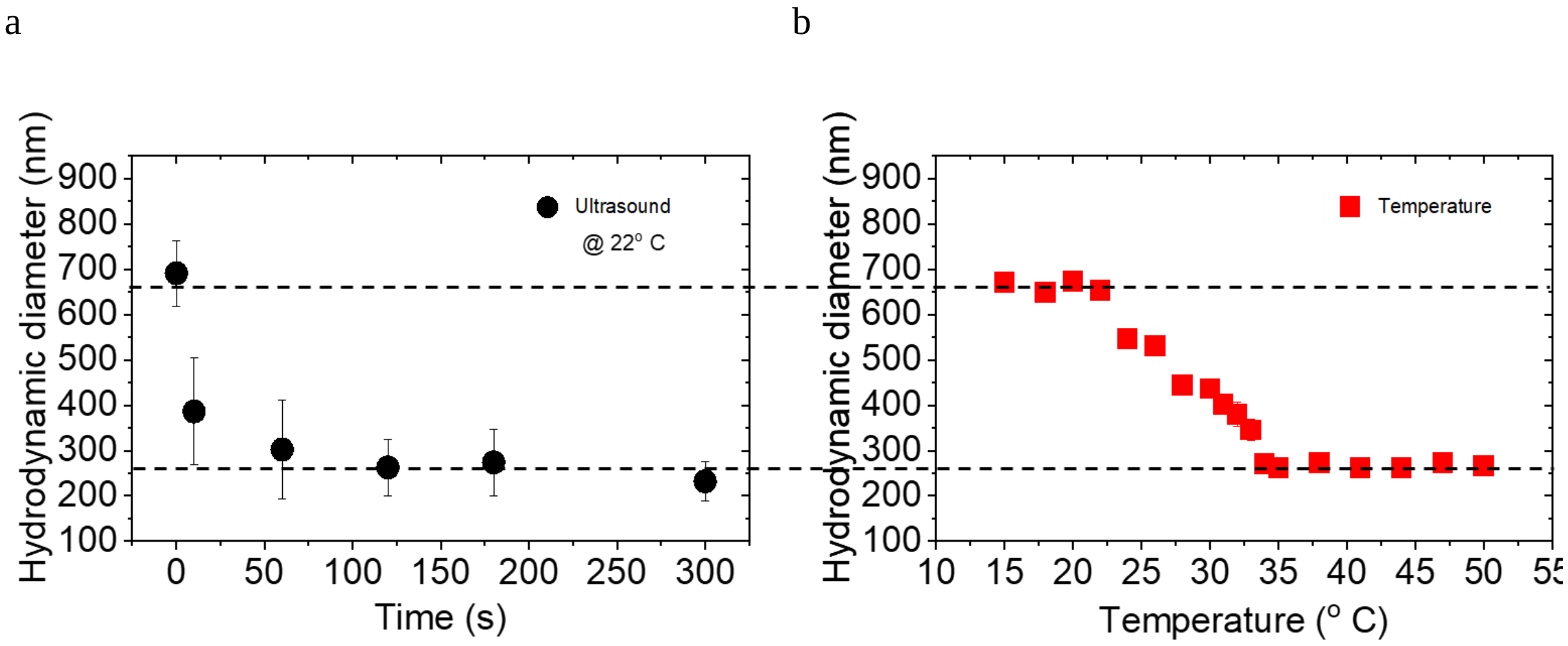

Figure 4. Hydrodynamic diameter of PNIPAM microgel (5 mol% cross-linker content), exhibiting the VPT due to both a) temperature change and b) ultrasonic actuation at room temperature (22° C). The US frequency and input voltage were set to 5.4 MHz and 400 mV respectively.

## V. Conclusion

In this work, we developed a DLS characterization of silica particles under the influence of US. In the theory part, we rewrite the continuity equation based on the time-dependent velocity of particles due to ultrasonic waves. Then, we derive a modified intensity autocorrelation function for dilute nano-spheres undergoing Brownian diffusion as well as ultrasonic vibration. The resulting model gives valuable information about the particle vibrational behavior in addition to the Brownian diffusion coefficient. The experimental work is performed using an US-DLS setup for silica particles of different sizes (from 80 nm to 1 μm) at different US frequencies (40 kHz to 5.4 MHz) and amplitudes. The findings with silica particles indicate that any potential disturbances, such as acoustic streaming, which could impact the size estimation of particles in DLS experiments, can be excluded. Therefore, the particle size can be correctly estimated even with the conventional fitting parameters for mean value of oscillating autocorrelation data points. Furthermore, we are able to extract the particle vibration amplitude and frequency in addition to their size successfully. This is because the frequency of particle vibration is high enough that the vibration and diffusion time-scales are decoupled. This method is important for the characterization of macromolecules and polymers whose size and behavior subjected to US are of interest. We used the developed setup to evaluate the shrinking behavior of PNIPAM microgels subjected to US. Our findings demonstrate that PNIPAM microgels are a notable example of acousto-responsive polymer networks. Their VPT in response to US holds a great potential for their application in drug delivery systems. In addition, US is a much faster trigger than temperature change.

**Acknowledgement**

Financial support from German Research Foundation (DFG)— Project Number (460540240) — is acknowledged.

**References**

[1] H. R. Schober and H. L. Peng, *Heterogeneous Diffusion, Viscosity, and the Stokes-Einstein Relation in Binary Liquids*, Phys. Rev. E **93**, 1 (2016).

[2] A. Einstein, *Investigations on the Theory Brownian Movement* (DOVER PUBLICATIONS, INC, 1956).

[3] S. Bhattacharjee, *DLS and Zeta Potential - What They Are and What They Are Not?*, J. Control. Release **235**, 337 (2016).

[4] K. Kang, *Image Time-Correlation, Dynamic Light Scattering, and Birefringence for the Study of the Response of Anisometric Colloids to External Fields*, Rev. Sci. Instrum. **82**, (2011).

[5] M. Mittal, P. P. Lele, E. W. Kaler, and E. M. Furst, *Polarization and Interactions of Colloidal Particles in Ac Electric Fields*, J. Chem. Phys. **129**, (2008).

[6] F. Mantegazza, M. Caggioni, M. L. Jiménez, and T. Bellini, *Anomalous Field-Induced Particle Orientation in Dilute Mixtures of Charged Rod-like and Spherical Colloids*, Nat. Phys. **1**, 103 (2005).

[7] Q. Li, J. Wang, J. Wang, J. Baleta, C. Min, and B. Sundén, *Effects of Gravity and Variable Thermal Properties on Nanofluid Convective Heat Transfer Using Connected and Unconnected Walls*, Energy Convers. Manag. **171**, 1440 (2018).

[8] R. Prasher, P. Bhattacharya, and P. E. Phelan, *Thermal Conductivity of Nanoscale Colloidal Solutions (Nanofluids)*, Phys. Rev. Lett. **94**, 3 (2005).

[9] X. Feng, G. Huang, J. Qiu, L. Peng, K. Luo, D. Liu, and P. Han, *Dynamic Light Scattering in Flowing Dispersion*, Opt. Commun. **531**, (2023).

[10] R. Weber and G. Schweiger, *Simultaneous In-Situ Measurement of Local Particle Size, Particle Concentration, and Velocity of Aerosols*, J. Colloid Interface Sci. **210**, 86 (1999).

[11] A. B. Leung, K. I. Suh, and R. R. Ansari, *Particle-Size and Velocity Measurements in Flowing Conditions Using Dynamic Light Scattering*, Appl. Opt. **45**, 2186 (2006).

[12] W. Krahn, M. Luckas, and K. Lucas, *Application of Photon Correlation Spectroscopy To Brownian Motion System.*, **23**, 813 (1986).

[13] T. W. Taylor and C. M. Sorensen, *Gaussian Beam Effects on the Photon Correlation Spectrum from a Flowing Brownian Motion System*, Appl. Opt. **25**, 2421 (1986).

[14] L. J. Delaney, S. Isguven, J. R. Eisenbrey, N. J. Hickok, and F. Forsberg, *Making Waves: How Ultrasound-Targeted Drug Delivery Is Changing Pharmaceutical Approaches*, Mater. Adv. **3**, 3023 (2022).

[15] W. Bonrath, *Ultrasound Supported Catalysis*, Ultrason. Sonochem. **12**, 103 (2005).

[16] B. Jin, H. Bang, and K. S. Suslick, *Applications of Ultrasound to the Synthesis of Nanostructured Materials*, 1039 (2010).

[17] A. Razavi, M. Rutsch, S. Wismath, M. Kupnik, R. von Klitzing, and A. Rahimzadeh,

*Frequency-Dependent Ultrasonic Stimulation of Poly(N-Isopropylacrylamide) Microgels in Water*, Gels **8**, 10, 628 (2022).

[18] A. Rahimzadeh, M. Rutsch, M. Kupnik, and R. von Klitzing, *Visualization of Acoustic Energy Absorption in Confined Aqueous Solutions by PNIPAM Microgels: Effects of Bulk Viscosity*, Langmuir **37**, 5854 (2021).

[19] B. J. Berne and R. Pecora, *Dynamic Light Scattering: With Applications to Chemistry, Biology, and Physics* (Dover Publications, 2000).

[20] O. Glatter, *Chapter 11 - Dynamic Light Scattering (DLS)*, in *Scattering Methods and Their Application in Colloid and Interface Science*, edited by O. B. T.-S. M. and their A. in C. and I. S. Glatter (Elsevier, 2018), pp. 223–263.

[21] S. Stock, S. Röhl, L. Mirau, M. Kraume, and R. von Klitzing, *Maximum Incorporation of Soft Microgel at Interfaces of Water in Oil Emulsion Droplets Stabilized by Solid Silica Spheres*, Nanomaterials **12**, 1 (2022).

[22] S. Stock, F. Jakob, S. Röhl, K. Gräff, M. Kühnhammer, N. Hondow, S. Micklethwaite, M. Kraume, and R. von Klitzing, *Exploring Water in Oil Emulsions Simultaneously Stabilized by Solid Hydrophobic Silica Nanospheres and Hydrophilic Soft PNIPAM Microgel*, Soft Matter **17**, 8258 (2021).

[23] C. E. Shannon, *Communication in the Presence of Noise*, Proc. IRE **37**, 10 (1949).

[24] A. Green, J. S. Marshall, D. Ma, and J. Wu, *Acoustic Streaming and Thermal Instability of Flow Generated by Ultrasound in a Cylindrical Container*, Phys. Fluids **28**, (2016).